\documentclass[10pt,aps,prl,twocolumn,superscriptaddress]{revtex4}
\usepackage[linktocpage=true, colorlinks=true, urlcolor=blue, linkcolor=blue, citecolor=blue]{hyperref}
\usepackage{graphicx}
\usepackage{amsmath,amssymb}
\usepackage{color}
\usepackage{url}
\usepackage{hyperref}
\usepackage{cleveref}
\usepackage{xfrac}
\usepackage{upgreek}
\usepackage{lipsum}
\usepackage{siunitx}

\graphicspath{{./Images/}}

\begin{document}

\title{Laser Doppler holography of the anterior segment for blood flow imaging, eye tracking, and transparency assessment}

\author{L\'eo Puyo}
\affiliation{Corresponding author: gl.puyo@gmail.com}
\affiliation{Centre Hospitalier National d'Ophtalmologie des Quinze-Vingts, INSERM-DHOS CIC 1423. 28 rue de Charenton, 75012 Paris, France}
\affiliation{Paris Eye Imaging, France}
\affiliation{Institute of Biomedical Optics, University of L\"ubeck. Peter-Monnik-Weg 4, 23562 L\"ubeck, Germany}

\author{Cl\'ementine David}
\affiliation{Centre Hospitalier National d'Ophtalmologie des Quinze-Vingts, INSERM-DHOS CIC 1423. 28 rue de Charenton, 75012 Paris, France}

\author{Rana Saad}
\affiliation{Centre Hospitalier National d'Ophtalmologie des Quinze-Vingts, INSERM-DHOS CIC 1423. 28 rue de Charenton, 75012 Paris, France}
\affiliation{Paris Eye Imaging, France}

\author{Sami Saad}
\affiliation{Centre Hospitalier National d'Ophtalmologie des Quinze-Vingts, INSERM-DHOS CIC 1423. 28 rue de Charenton, 75012 Paris, France}

\author{Josselin Gautier}
\affiliation{Centre Hospitalier National d'Ophtalmologie des Quinze-Vingts, INSERM-DHOS CIC 1423. 28 rue de Charenton, 75012 Paris, France}
\affiliation{Paris Eye Imaging, France}

\author{Jos\'e-Alain Sahel}
\affiliation{Centre Hospitalier National d'Ophtalmologie des Quinze-Vingts, INSERM-DHOS CIC 1423. 28 rue de Charenton, 75012 Paris, France}
\affiliation{Paris Eye Imaging, France}
\affiliation{Department of Ophthalmology, The University of Pittsburgh School of Medicine. Pittsburgh, PA, 15213, United States}
\affiliation{Institut de la Vision. Sorbonne Universit\'e. INSERM, CNRS. 17 Rue Moreau, 75012 Paris, France}

\author{Vincent Borderie}
\affiliation{Centre Hospitalier National d'Ophtalmologie des Quinze-Vingts, INSERM-DHOS CIC 1423. 28 rue de Charenton, 75012 Paris, France}

\author{Michel Paques}
\affiliation{Centre Hospitalier National d'Ophtalmologie des Quinze-Vingts, INSERM-DHOS CIC 1423. 28 rue de Charenton, 75012 Paris, France}
\affiliation{Paris Eye Imaging, France}
\affiliation{Institut de la Vision. Sorbonne Universit\'e. INSERM, CNRS. 17 Rue Moreau, 75012 Paris, France}

\author{Michael Atlan}
\affiliation{Paris Eye Imaging, France}
\affiliation{Institut Langevin. Centre National de la Recherche Scientifique (CNRS). Paris Sciences \& Lettres (PSL University). \'Ecole Sup\'erieure de Physique et de Chimie Industrielles (ESPCI Paris) - 1 rue Jussieu. 75005 Paris France}

\date{\today}

\begin{abstract}
Laser Doppler holography (LDH) is a full-field blood flow imaging technique able to reveal human retinal and choroidal blood flow with high temporal resolution. We here report on using LDH in the anterior segment of the eye without making changes to the instrument. Blood flow in the bulbar conjunctiva and episclera as well as in corneal neovascularization can be effectively imaged. We additionally demonstrate simultaneous holographic imaging of the anterior and posterior segments by simply adapting the numerical propagation distance to the plane of interest. We used this feature to track the movements of the retina and pupil with high temporal resolution. Finally, we show that the light backscattered by the retina can be used for retro-illumination of the anterior segment. Hence digital holography can reveal opacities caused by absorption or diffusion in the cornea and eye lens.
\end{abstract}

\maketitle

\section{Introduction}

Laser Doppler holography (LDH) is a digital holographic method where blood flow is measured from the interference between light backscattered by the eye and a reference beam~\cite{Pellizzari2016}. The Doppler broadening is measured over the full-field array of a camera, and thanks to the coherent gain brought by the reference beam, ultrafast frame rates can be used to record optical field fluctuations up to a few tens of kHz. A short-time Fourier transform analysis can be used to measure the variations of human retinal blood flow over cardiac cycles with a few milliseconds of temporal resolution~\cite{Puyo2018}, allowing to measure the resistivity index in all the arteries of the field of view~\cite{Puyo2019b}. In our previous work, we used LDH to image the choroidal vasculature~\cite{Puyo2019}, introduced a digital method to improve the contrast of low frequency power Doppler images by singular value decomposition filtering~\cite{Puyo2020}, and presented a way to perform LDH measurements with lower sampling frequencies~\cite{Puyo2020b}.

We here further explored the capabilities of LDH to image the anterior segment of the eye and track the eye movements. Cataract is a leading cause of vision loss that affects mostly the elderly population~\cite{Pascolini2012, Flaxman2017}. To this day, in clinical practice the decision to operate the cataract remains based on the subjective appreciation of the apparent opacity of the eye lens observed with slit-lamp microscopy. The optical quality impairment caused by cataract has become one of the major indications for cataract surgery~\cite{lundstrom2015changing}. The management of cataract might thus benefit from objective assessments of the eye transparency. Other techniques have been proposed to characterize the lens transparency, based on the imaging of the crystalline lens~\cite{Chylack1993, Kirkwood2009}, or on the degradation of the retinal image quality~\cite{Van2007, Bueno2007, Galliot2016, hwang2018utility}. A technical difficulty to overcome is to separate the contributions of refraction errors from those due of diffusion and absorption, and to give a grade that correlates as much as possible with the subject vision.
Another recurrent issue in the anterior segment is corneal neovascularization, which is caused by a disruption in the balance of angiogenic and antiangiogenic factors, often secondary to inflammation or hypoxia~\cite{Chang2001corneal}. Left untreated, corneal neovascularizations can lead to persistent inflammation and scarring, and ultimately threaten the corneal transparency and visual acuity~\cite{abdelfattah2015clinical}. However, treatment of corneal neovascularization is a complex issue~\cite{feizi2017therapeutic}. In clinical practice, the angiography of corneal neovascularization can be realized with fluorescent dyes, or with optical coherence tomography (OCT)~\cite{brunner2018imaging, steger2016corneal, palme2018functional}. However, these methods do not offer a quantitative blood flow contrast, and therefore provide mainly structural information about the neovessels. Blood flow imaging could thus be helpful to identify the most suitable therapeutic approach by providing better means to characterize corneal neovasculature, and to better follow-up the effect of treatment. Blood flow imaging in the anterior segment could also prove valuable to study other ocular surface vasculatures such as the conjunctiva, the episclera. Underlying clinical interests include the study of dry eyes, lesions, allergies, ulcers, effect of pollution, and viral and bacterial inflammation~\cite{la2013allergic, azari2013conjunctivitis}.
Finally, the imaging of the anterior segment is also valuable to perform a comprehensive monitoring of the eye movements, for both the understanding of oculomotor control and the resulting clinical applications. Eye movements are part of the vision process, and are controlled by several interacting regions of the central nervous system~\cite{leigh2015neurology}. Abnormal eye movements can be an indicator of neurodegeneration~\cite{anderson2013eye}, and eye movements dysfunctions have also been reported in patients with psychiatric diseases~\cite{trillenberg2004eye}. Although serious alterations of oculomotor control can be detected during clinical examination, the detection of slight abnormalities of pursuit, saccadic, optokinetic, and vestibulo-ocular movements requires a quantitative monitoring with a good spatio-temporal resolution~\cite{danchaivijitr2004diplopia}. For example, while microsaccades in controls are generally horizontal, obliquely oriented microsaccades have been reported to be more common in mild moderate Alzheimer's and to occur at a more elevated rate~\cite{molitor2015eye}. Ultimately, monitoring the eye movements might prove helpful to assess the efficiency of neuro-protective and neuro-restorative therapies by following-up the severity of symptoms, and might also aid clinicians establish the diagnosis of neurodegenerative disorders~\cite{macaskill2016eye}.

We report here on exploring the potential clinical uses of LDH to address these issues. We used the exact same instrument as the one previously used to image retinal and choroidal blood flow. The interferograms were merely numerically propagated to have the anterior segment in focus instead of the eye fundus. We explored the possibility to image blood flow in the surface vasculatures of the anterior segment, in the conjunctiva/episclera as well as in cases of corneal neovascularization. We then demonstrate the tracking of the movements of the anterior and posterior segments of the eye from the same LDH data set. Finally, we imaged the eye of subjects affected by cataract or corneal scarring and showed that holographic reconstructions in the pupil plane of light backscattered by the retina can reveal the absorbing or diffusing features in the cornea and lens.

\section{Methods}


We used the fiber Mach-Zehnder LDH setup previously developed to image retinal and choroidal blood flow in human subjects and presented in~\cite{Puyo2018, Puyo2019}. We either used as a light source for the experiments a $\SI{50}{\milli\watt}$ single-frequency laser diode operating at $\SI{785}{\nano\meter}$ wavelength (Thorlabs LP785-SAV50, VHG wavelength-stabilized, coherence length $\SI{20}{\meter}$), and a $\SI{30}{\milli\watt}$ single frequency laser diode operating at $\SI{852}{\nano\meter}$ (Thorlabs FPV852S, VHG wavelength-stabilized, coherence length $\SI{30}{\meter}$). The eye was exposed to $\SI{2}{\milli\watt}$ or less of constant exposure focused in front of the eye, at a distance either equal to or greater than the eye focal length, depending on the desired field of view. When focusing exactly in the eye front focal plane, the obtained field of view has a similar extension on the pupil and retina. When setting the eye backward, the laser beam diverges more before it reaches the eye and the field of view on the anterior segment is increased whereas the anterior segment irradiance is lower. At the same time the illuminated retinal area is reduced as the eye refraction is greater, but the amount of power received by the retina is also dramatically decreased so that it stays largely in check with safety limitations. For these experiments we considered the exposure limitations specified by ISO 15004-2:2007 for non-hazardous ophthalmic instruments (i.e. group 1), according to which, in this situation the irradiance of the anterior segment should not exceed $\SI[inter-unit-product = \ensuremath{{}\cdot{}}]{20}{\milli\watt\per\cm\squared}$ (unweighted by the wavelength) on a disc area of 1 mm of diameter. This limit on the anterior segment is far more limiting than the retinal maximum permissible exposure according to this same standard, which is $\SI[inter-unit-product = \ensuremath{{}\cdot{}}]{1.0}{\watt\per\cm\squared}$ at $\SI{785}{\nano\meter}$, and $\SI[inter-unit-product = \ensuremath{{}\cdot{}}]{1.4}{\watt\per\cm\squared}$ at $\SI{850}{\nano\meter}$. Experimental procedures adhered to the tenets of the Declaration of Helsinki, and the study was approved by an ethical committee (Comit\'e de Protection des Personnes; clinical trial NCT04129021). Written informed consent was obtained from all subjects.

The light backscattered by the eye interferes with an on-axis reference beam to form holograms digitally recorded by an ultrafast CMOS camera (Ametek - Phantom V2511, 12-bit pixel depth, pixel size $\SI{28}{\micro\meter}$) in a $512 \times 512$ format, with a sampling frequency $f_{\rm S}$ that here varied between 8 and 67 kHz. The exposure time was usually set to the maximum possible depending on the sampling rate in order to fully benefit from the light backscattered by the eye. The data processing remains the same as for retinal and choroidal blood flow imaging. In essence, a short-time Fourier transform analysis with a sliding window duration varying between 1 and 30 ms is carried out to reveal local blood flow variations from the pulsatile changes of Doppler broadening. Each interferogram recorded by the camera is numerically propagated to the plane of interest by angular spectrum propagation~\cite{Goodman2005, Puyo2018}. The short-time window is then filtered by singular value decomposition (SVD) to remove the Doppler contribution of eye motion~\cite{Puyo2020}. The eigenvectors of the holograms space-time matrix associated to the eigenvalues of highest energy are rejected to allow the access to blood flow contributions of low frequency. The number of rejected eigenvectors is always set accordingly with the lower frequency threshold used to compute the power Doppler~\cite{Puyo2020}. The Doppler power spectrum density is then computed from the squared magnitude of the temporal Fourier transform of the holograms short-time window~\cite{Puyo2018}, and it is integrated between arbitrarily chosen frequencies. Finally, a compensation is made for the field vignetting, and the baseline signal is subtracted~\cite{Puyo2019b}.


\section{Blood flow imaging}

\begin{figure}[t!]
\centering
\includegraphics[width = 1\linewidth]{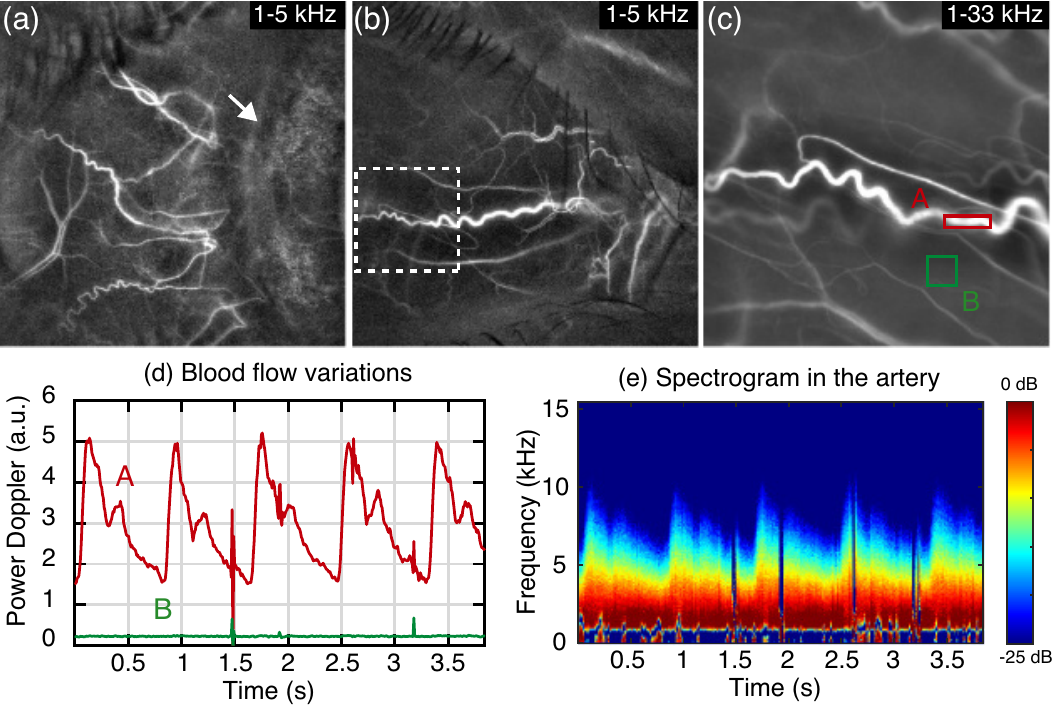}
\caption{Blood flow imaging in episcleral and conjunctival vessels. (a-c) Power Doppler images, the arrow points to the junction between the conjunctiva and the iris. (d) Power Doppler variations in the indicated ROIs. (e) Spectrogram measured in the artery. \textcolor{blue}{\href{https://youtu.be/okeKJWfpP3Y}{Visualization 1}}.
}
\label{fig_1_FieldOfView}
\end{figure}

In Fig.~\ref{fig_1_FieldOfView} are shown examples of LDH blood flow images in the anterior segment of healthy subjects. In Fig.~\ref{fig_1_FieldOfView}(a), the measurement was performed on the eye temporal side at a 10 kHz frame rate. Vessels from the episclera and bulbar conjunctiva are revealed, and the shadow of the upper eyelash is visible. The size of the smallest visible vessels is estimated to about 10-$\SI{20}{\micro\meter}$. Vessels in the iris are not revealed here (the arrow indicates the outline of the iris). This may be because the iris had a dark pigmentation. It is also possible that the flow velocity in iris vessels is slower, so that they would be harder to reveal with LDH as lower frequencies are not totally free of bulk motion contributions that decrease the contrast of blood flow images. In Fig.~\ref{fig_1_FieldOfView}(b), the measurement was performed on the nasal side of the pupil, also at a 10 kHz frame rate. Once again, vessels from the episclera and bulbar conjunctiva are revealed. The shadows of the upper and lower eyelashes are visible, and even a cutaneous vessel can be observed in the upper eyelid. In \textcolor{blue}{\href{https://youtu.be/okeKJWfpP3Y}{Visualization 1}} is shown the corresponding blood flow movie where the eye is laterally shifted during the measurement to explore a greater area.
In Fig.~\ref{fig_1_FieldOfView}(c) is shown the power Doppler image from a measurement performed over a smaller field of view in the same eye (dashed square), and blood flow variations are monitored with power Doppler and spectrograms. The measurement was conducted at 67 kHz in order to characterize the Doppler spectrum characteristics in the episclera/conjunctiva up to elevated frequencies. In Fig.~\ref{fig_1_FieldOfView}(d), are shown the variations of power Doppler in the regions of interest drawn in Fig.~\ref{fig_1_FieldOfView}(c) marking an artery (red) and the background (green). Blood flow in the artery shows a systolic peak, dicrotic notch, and a progressive diastolic blood flow decrease, which looks like the typical waveform that can be observed in retinal arteries with LDH~\cite{Puyo2019b}. The blood flow in this artery generates a strong and reproducible Doppler signal monitored over consecutive cardiac cycles.
The quality of the blood flow measurement is sufficient to characterize the blood flow variations and thus the resistivity index, calculated as ${\rm RI} = \frac{V_{\rm systole} - V_{\rm diastole}}{V_{\rm systole}}$, and supposed to reflect the resistance of the perfused vascular bed ($V$ being the flow velocity).
Finally, the spectrogram in the same arterial region of interest in Fig.~\ref{fig_1_FieldOfView}(e) shows the spectral distribution of energy. Despite this vessel being of large size compared to the other vessels of conjunctiva/episclera present in the field of view, there is no energy in the spectrum above 8-10 kHz. This stands in contrast with retinal vessels where Doppler shifts usually reach a few tens of kHz. The direct consequence is that a lower sampling frequency is required to image blood flow in the conjunctiva/episclera. However, this also makes conjunctival/episcleral blood flow imaging more challenging because the SVD rejection of Doppler contribution of bulk motion is not perfect so low frequency Doppler responses from blood flow are more vulnerable to eye motion.

\begin{figure}[t!]
\centering
\includegraphics[width = 1\linewidth]{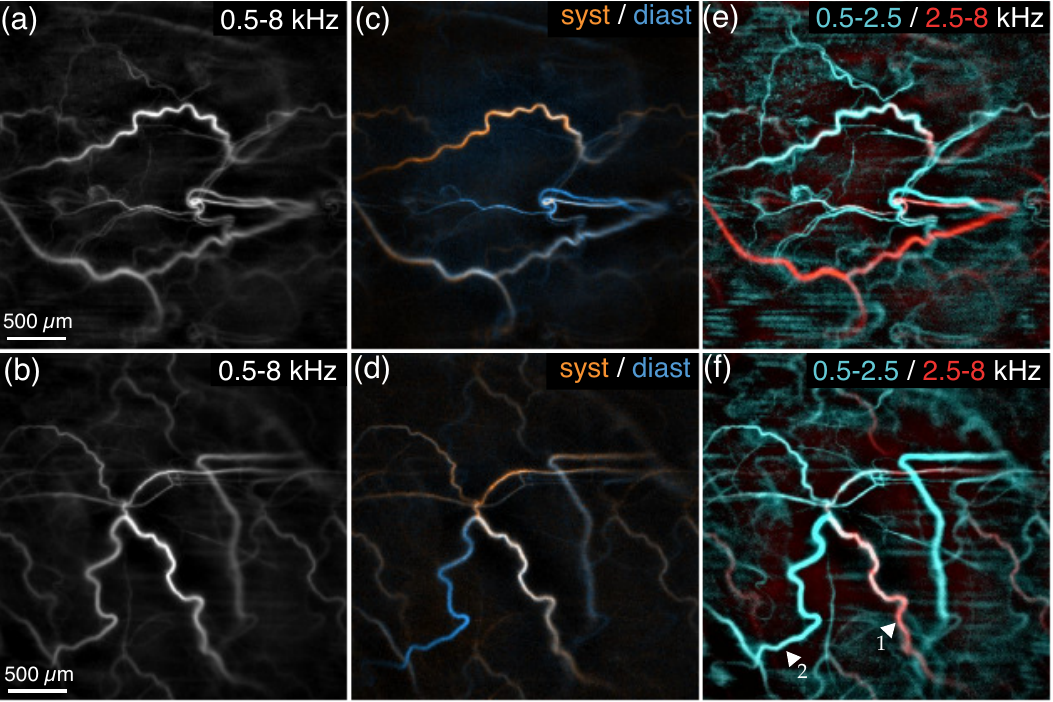}
\caption{Functional contrast in the conjunctiva/episclera with LDH. (a-b) Power Doppler images. (c-d) Composite images of systole/diastole in orange/blue. (e-f) Composite image of slow/fast flows in cyan/red. Corresponding color composite power Doppler movies in \textcolor{blue}{\href{https://youtu.be/3oA-4cDaC8E}{Visualization 2}} and \textcolor{blue}{\href{https://youtu.be/3uihOLdLuMA}{Visualization 3}}.
}
\label{fig_2_PD_SD_cmp}
\end{figure}

In Fig.~\ref{fig_2_PD_SD_cmp} are presented two examples of LDH images of the anterior segment with blood flow contrasts that depend on the flow velocity and on the blood flow changes occurring throughout cardiac cycles. The two measurements were performed at 67 kHz, but only the frequencies up to 8 kHz were used for these images, as we found that frequencies above 8 kHz contained more noise than actual blood flow signal, and therefore that using the very high frequencies was detrimental to the images signal-to-noise ratio. Standard power Doppler images are shown in Fig.~\ref{fig_2_PD_SD_cmp}(a) and (b) from the frequency range 0.5-8 kHz. In Fig.~\ref{fig_2_PD_SD_cmp}(c) and (d) are shown composite images obtained by merging power Doppler images averaged during systole (orange) and diastole (blue). In the retina, thanks to the differences of flow variations between in arteries and veins, this process allows identifying the vessels' type~\cite{Puyo2019b, Puyo2020}.
In the systole/diastole images shown here, a differentiation of vessels is also visible, and it is likely due to the difference of flow variations between arteries and veins. However, it cannot be excluded at this point that this difference of contrast is not due to the fact that vessels belong to different vasculatures (conjunctiva/episclera). Finally, in Fig.~\ref{fig_2_PD_SD_cmp}(e) and (f), are shown colored composite images obtained by merging low and high frequency power Doppler images in cyan and red, respectively~\cite{Puyo2019}. In the choroid, such a process helps distinguishing the arteries from the veins. A differentiation of vessels is also visible here, and once again it could be thought that it is episcleral and conjunctival vessels that are differentiated. It seems more likely based on the differences of flow velocity between arteries and veins, similarly to what is observed in the choroid with LDH. For example, based on their disposition and size, it seems realistic to assume that the vessels '1' and '2' belong to the same vasculature, and that '1' is an artery and '2' is a vein, since vessel '1' carries a faster flow than vessel '2'.


\section{Eye movement monitoring}

\begin{figure}[t!]
\centering
\includegraphics[width = 1\linewidth]{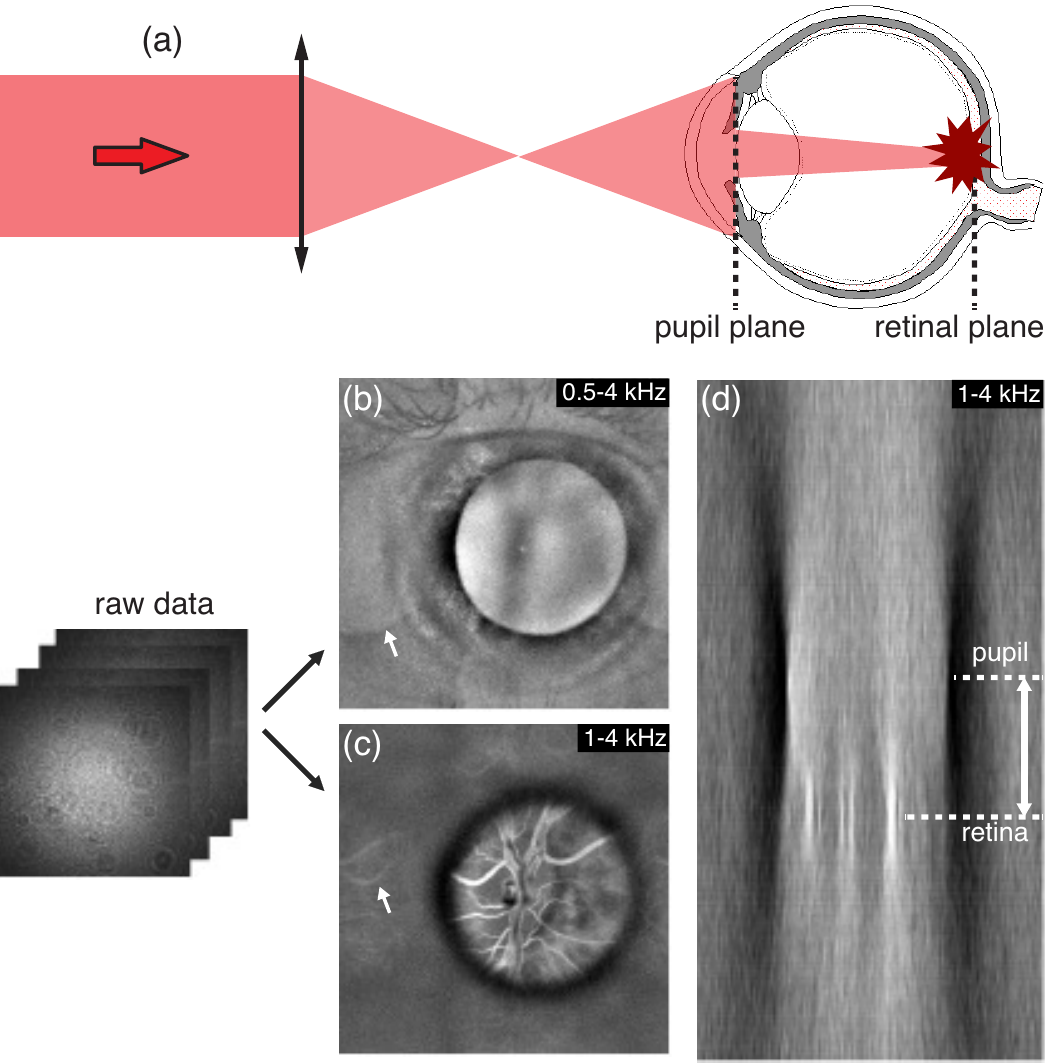}
\caption{Simultaneous holographic imaging of the anterior and posterior segments of the eye. (a) The whole eye is illuminated, and the numerical reconstruction is performed to have in focus either (b) the pupil plane, or (c) the retinal plane. (d) Cross-sectional holographic view. \textcolor{blue}{\href{https://youtu.be/xucjbZ4wSow}{Visualization 4}} illustrates a dynamic refocusing between the camera, pupil, and retinal planes.
}
\label{fig_4_AnteriorPosteriorSegment}
\end{figure}

In the optical configuration we used, the laser beam illuminates over an extended area both the anterior and posterior segments of the eye. The coherence length of the single frequency laser diode is largely sufficient to allow light backscattered from both parts of the eye to interfere with the reference field on the camera sensor. It is therefore possible to reconstruct the holographic images of a given dataset in either the anterior or the posterior segment of the eye, as illustrated in Fig.~\ref{fig_4_AnteriorPosteriorSegment} with an 8 kHz measurement. The angular spectrum propagation was performed to compute power Doppler images respectively in the pupil and retinal plane in Fig.~\ref{fig_4_AnteriorPosteriorSegment}(b) and (c). The eye of the subject was set back to image a very large field of view in the pupil plane, seemingly about 15 mm wide. Due to the refractive power of the eye, the extent of the retinal field of view is smaller than the actual pupil, and can be estimated to approximately 5 mm wide based on the extents of the retinal vessels and optic nerve head (although obstructed by the pupil).
The reconstruction distances were respectively 16 cm and 34 cm for the pupil and retinal planes, which significantly blurs the structures from the other plane. The retinal vessels in the anterior segment reconstruction are significantly but not completely defocused, as their blurred presence can be seen in the pupil area with a dark contrast due to the low frequency range~\cite{Puyo2020b}.
Outside the pupil area, no vessels can be seen in the anterior segment, probably because the laser energy is spread over a very large field and the collection of backscattered light is not optimized. Finally, in Fig.~\ref{fig_4_AnteriorPosteriorSegment}(d) is shown a cross-sectional view of the eye. It was obtained by computing the numerical propagation and power Doppler analysis for different reconstruction distances. The bright structures in focus in the retinal plane are blood vessels, whereas the pupil plane is recognized as the plane in which the contour of the pupil comes into focus.

When looking carefully at the blood flow image reconstructed in the retinal plane in Fig.~\ref{fig_4_AnteriorPosteriorSegment}(c), the presence of vascular structures outside the pupil area can be noticed (arrow). In the anterior segment reconstruction in Fig.~\ref{fig_4_AnteriorPosteriorSegment}(b), the pupil is also laterally replicated (arrow). These replicas are artifacts caused by spatially aliased interferences. The angular acceptance cone of the camera pixels gives the limitation for the numerical aperture that can be used. In our case the camera pixels are quite large ($\SI{28}{\micro\meter}$), which significantly limits the numerical aperture that can be used (the exact value depends on the optical magnification in use). Part of the object light that interferes with the reference beam on the camera generates fringes on the sensor array that are of too high frequency relatively to the pixel size. The undersampled speckles lead to the apparition of replicas of the main holographic image regularly positioned in space~\cite{Allebach1976, kelly2008practical}. In standard holographic imaging, the spatial frequency content of the object beam impinging the camera can be optically limited by placing an iris in a Fourier conjugate plane, or by adapting the object-camera distance so that the camera acts as an aperture stop~\cite{Hillmann2011}. The replicas are removed once the frequency content of the interference between the incoming object light and the reference beam fits into the camera bandwidth. In the configuration we used, no physical aperture stop was used in the reciprocal plane of the anterior segment. In order to avoid having the replicas overlap with the normal holographic image, we moved the subject eyes backwards for these experiments. Interestingly, on the second imaging channel used for real-time monitoring~\cite{Puyo2018}, the replicas were not seen on the camera with $\SI{12}{\micro\meter}$ large pixels (data not shown).

\begin{figure}[t!]
\centering
\includegraphics[width = 1\linewidth]{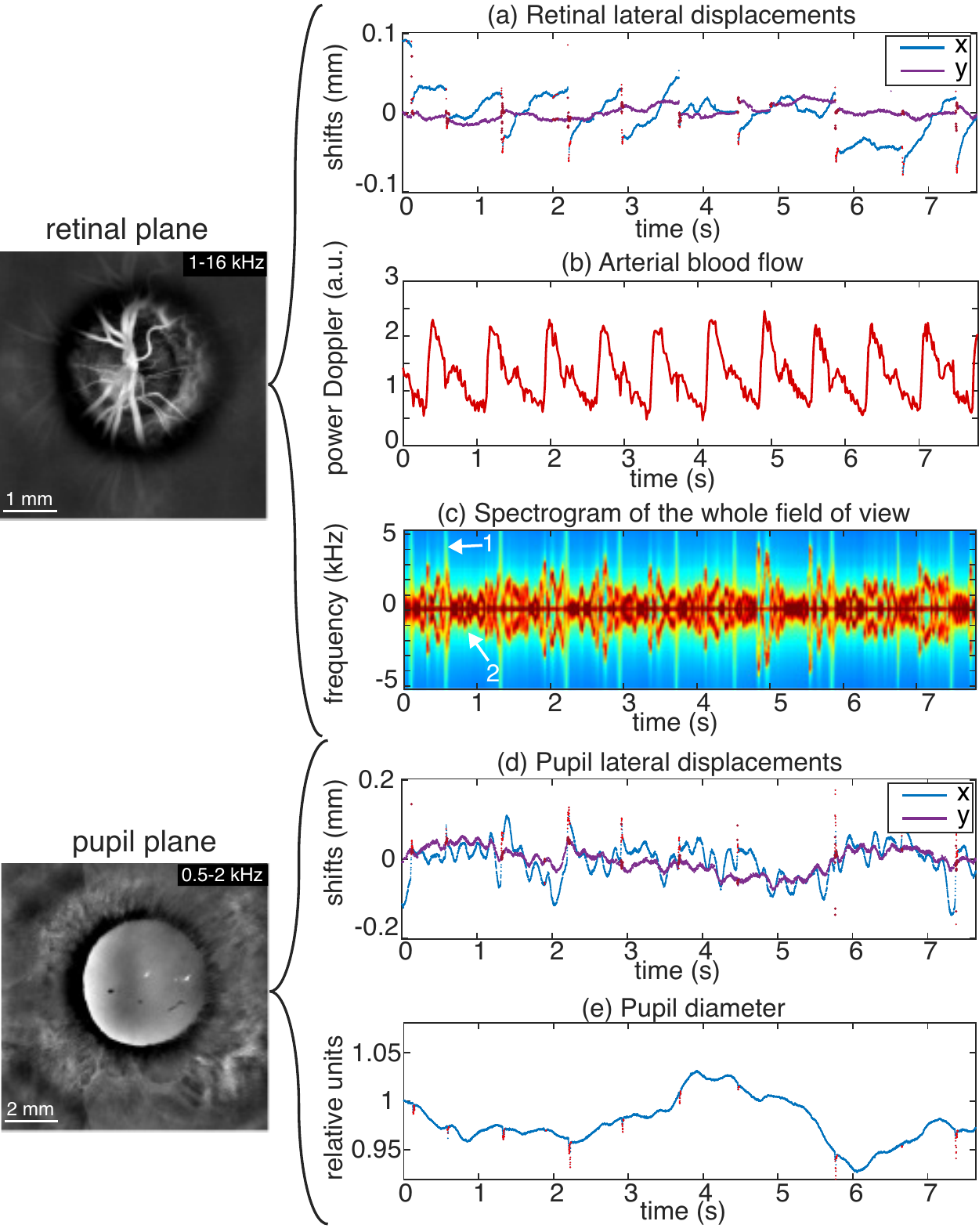}
\caption{Tracking the movements of an eye slightly affected strabismus with LDH. In the retinal plane are measured (a) the fundus lateral displacements, (b) the retinal arterial blood flow, and (c) the averaged Doppler spectrogram of the fundus revealing the signatures of in-plane saccades and out-of-plane motion. In the anterior segment reconstruction are tracked (d) the pupil lateral displacements, and (e) the pupil diameter. \textcolor{blue}{\href{https://youtu.be/iYBZ-tiSG_I}{Visualization 5}} shows the 6-16 kHz power Doppler movies in the retinal and pupil planes.
}
\label{fig_4_EyeMovements}
\end{figure}

This ability to simultaneously image the anterior and posterior segments of the eye can be used to track the eye movements with a high temporal resolution, as demonstrated in Fig.~\ref{fig_4_EyeMovements}. A LDH measurement was performed at a frame rate of 33 kHz with continuous camera recording, and the subject eye was positioned so that both the retina and pupil were imaged over an area sufficiently large to allow a robust registration in both planes. In Fig.~\ref{fig_4_EyeMovements}(a) are shown the lateral displacements of the retina that were measured from a power Doppler movie reconstructed with a temporal resolution of 1 ms (short-time window of 32 images). $x$ and $y$ are the horizontal and vertical axis in the retinal and pupil planes. The points in red are those where the correlation coefficient of the registered image with the reference frame was lower than an arbitrarily chosen threshold. One can see that these points occur when there is a discontinuity of retinal position, which shows that they correspond to micro-saccades, which degrade the blood flow contrast of power Doppler images due to the bulk motion. The second graph in Fig.~\ref{fig_4_EyeMovements}(b) shows the monitoring of arterial blood flow obtained from the variations of power Doppler measured in a retinal artery, with a temporal resolution of 31 ms (short-time window of 1024 images).
Finally, the spectrogram in Fig.~\ref{fig_4_EyeMovements}(c) brings complementary information about the eye movement. Two types of Doppler contributions can be distinguished based on their spectral distribution of energy, we assume they correspond to in-plane and out-of-plane bulk movements. The arrow '1' points to a spectral trace which can be identified as a micro-saccade because it occurs when there is a discontinuity in the sampled retinal trajectory. This spectral trace spreads over a large range of frequencies, which is consistent with the fact that micro-saccades cause an in-plane motion of the fundus. In-plane movements induce minimal Doppler frequency shifts to directly backscattered light. However, for multiply scattered light, the variety of scattering directions with respect to the direction of the micro-saccade movement induces a diversity of Doppler frequency shifts, which leads to a spread of the spectral energy distribution. On the contrary, the arrow '2' points to a red trace where the spectral energy is very localized, that should correspond to the out-of-plane bulk motion. The eye axial motion has been hypothesized to be caused by the choroidal pulsatile swelling or by heartbeat related movement of the whole head~\cite{Singh2010, Singh2011, DeKinkelder2011}. In that case, because the movement occurs along the optical axis, the backscattered light component is more Doppler shifted than multiply-scattered light for which scattering directions are randomized. As a consequence, since all backscattered light receives the same Doppler shift, the spectral energy distribution is very localized. Furthermore, this component of the spectrogram, averaged over the whole fundus, may reasonably be expected to scale linearly with the instantaneous out-of plane velocity of the retina.


The pupil contour blocks the Doppler shifted light backscattered from the retina, so it can be imaged with a similar signal-to-noise ratio as the retina. This can appreciated in \textcolor{blue}{\href{https://youtu.be/iYBZ-tiSG_I}{Visualization 5}}, where the 6-16 kHz power Doppler movies in the retinal and pupil planes are shown. Therefore, despite not detecting any blood flow structures in these wide-field anterior segment images, the size and position of the pupil can be monitored with the same temporal resolution as retinal displacements. This is illustrated in Fig.~\ref{fig_4_EyeMovements}(d). The lateral displacements of the pupil can be tracked thanks to the output of the registration of the pupil plane power Doppler movie. Finally, in Fig.~\ref{fig_4_EyeMovements}(e) is shown the measurement of the dynamic dilation and constriction of the pupil that can be jointly monitored with a temporal resolution of 1 ms.
Interestingly, the parameters that are monitored here throughout cardiac cycles are all uncorrelated to one another. The subject eye was affected by strabismus, which is presumably the reason for the particular fixational patterns in Fig.~\ref{fig_4_EyeMovements}(a), as the retina was laterally drifting and the subject corrected his direction of gaze every second or so. The subject's strabismus is probably also the reason for the intriguing oscillating pattern for the pupil lateral displacements.


\section{Lens and corneal transparency imaging}

\subsection{Cataract imaging}

\begin{figure}[t!]
\centering
\includegraphics[width = 1\linewidth]{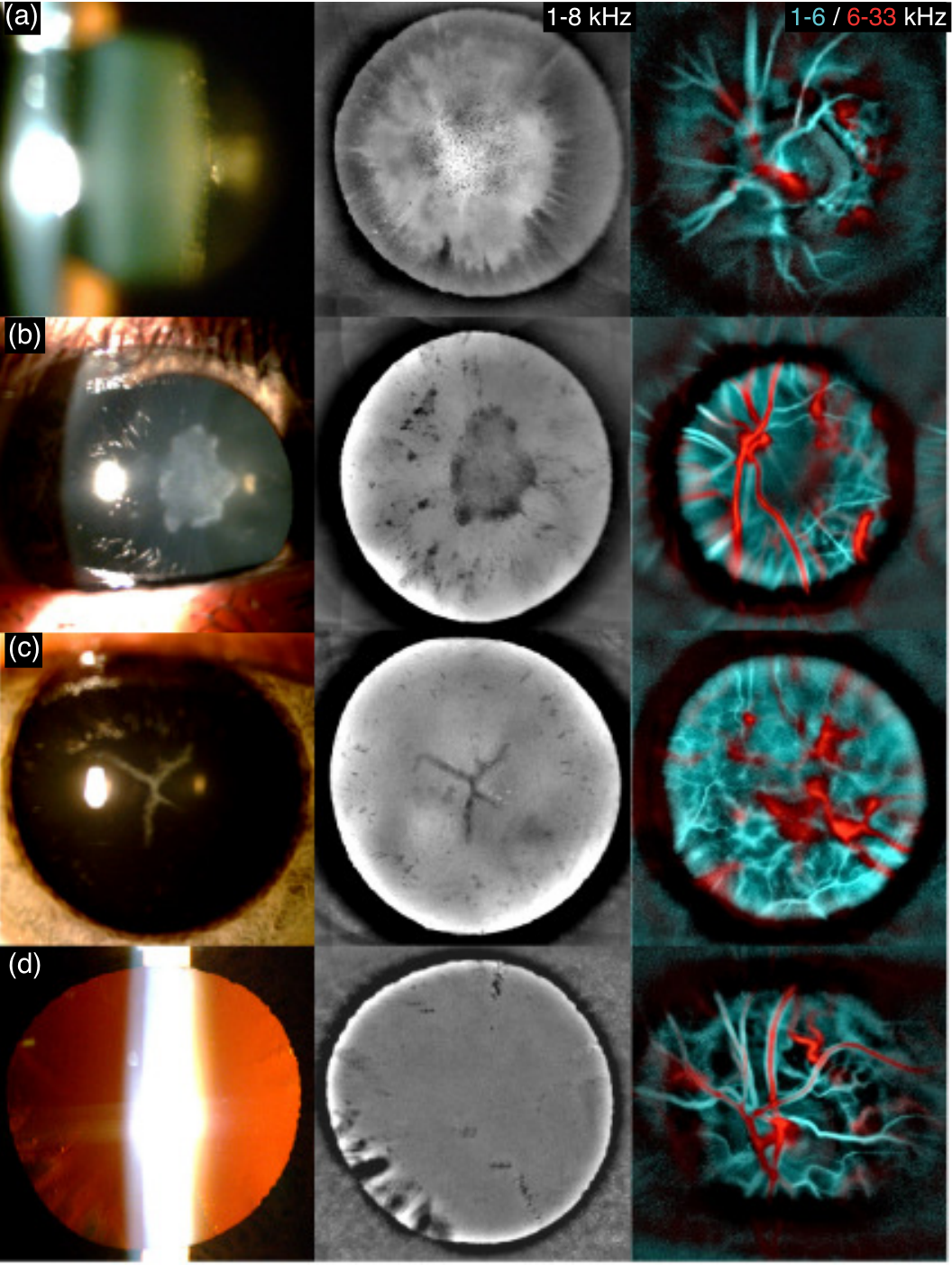}
\caption{Imaging crystalline lens opacification with LDH. For each eye are shown the slit lamp examination (left), and from the same LDH data we computed the 1-8 kHz power Doppler image in the anterior segment (middle), and the slow/fast flow color composite image in the posterior segment (right).
(a) Posterior subcapsular cataract.
(b) Anterior polar cataract.
(c) Star-shaped congenital cataract.
(d) Presence of riders.
}
\label{fig_5_Cataracte}
\end{figure}

Power Doppler images of the anterior segment show the pupil area thanks to the defocused light from the retina. Because these images are formed with the SVD and Fourier analyses that aim at revealing blood flow, the optical field in the anterior segment comes from the defocused fundus vasculature. Depending on the imaged area and the power Doppler frequency range, that can include light scattered by retinal large vessels, unresolved capillaries, or in the choroid. The Doppler shifted light from the retina performs a retro-illumination of the anterior segment through the pupil, which can be used to reveal the transparency of the refractive elements of the eye. As the light from the retina goes through the lens and cornea, the transmitted optical field is modulated by the structures that absorb or scatter the light. These opacities can be revealed when computing power Doppler images from fundus holograms reconstructed in the pupil plane.
This is illustrated in Fig.~\ref{fig_5_Cataracte} where four eyes affected by different forms of cataract are documented with slit lamp images (left column) and LDH. From the same LDH data, we computed the 1-8 kHz power Doppler image in the anterior segment (middle column), and the color composite image of slow/fast flows in the posterior segment (right column).

The first example shows an eye affected by posterior subcapsular cataract, the second is an eye affected by anterior polar cataract, the third is a case of star-shaped congenital cataract, and finally in the last example there were riders (opacifications of the border of the lens oriented towards the lens center). In all examples LDH offers a better contrast than the slit lamp to delineate the opacities, which allows resolving smaller non-transparent features. LDH also presents the advantage of being able to image the eye transparency over the full extent of the dilated pupil without suffering from any specular reflection. Indeed, when imaging the anterior segment, ophthalmic instruments may suffer from strong corneal specular reflections. For example, slit lamp images are obtained with an oblique illumination to reduce specular reflection but they are still significantly affected by parasitic reflections that hide the contrast of the structures below them. However, LDH is able to bring a good contrast to the structures causing absorption of light over the full pupil without having any issue due to specular reflections thanks to optical and digital filtering. A polarizing beamsplitter cube is used to illuminate the eye with a linear polarization and only the cross-polarized light backscattered by the eye is transmitted to the camera. Therefore, the corneal specular reflection which preserves the initial polarization is optically rejected. Secondly, the Fourier filtering also ensures that the reconstructed light field is due to Doppler shifted light, which in the pupil area is the defocused light from the fundus. Indeed, the specular reflection should only be slightly Doppler shifted, and in case it is Doppler shifted due to the eye's axial motion, then it should be rejected by the SVD filtering due to the high spatial coherence of the Doppler shift~\cite{Puyo2020}.

\begin{figure}[t!]
\centering
\includegraphics[width = 1\linewidth]{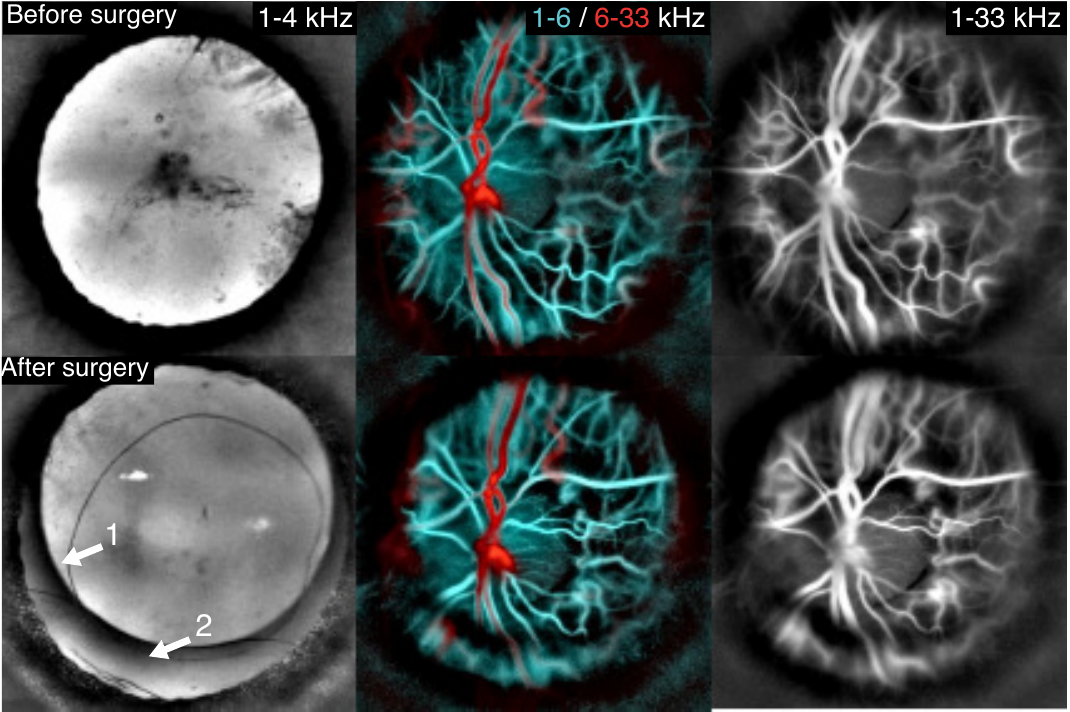}
\caption{Imaging an eye affected by cataract before and after surgery. Before the operation, opacities are present in the center and periphery of the eye lens, but fundus imaging with LDH can still be performed. After the operation, the mark of the implant is visible in the anterior segment, and a similar quality of imaging is obtained in the fundus.
}
\label{fig_7_surgery}
\end{figure}

In LDH pupil images shown in Fig.~\ref{fig_5_Cataracte}, no defocused retinal vasculature is seen, and there are probably several reasons for that. First, we used frequency ranges more favorable to having a more homogeneous illumination from the fundus. If using the 6-33 kHz power Doppler range (red part of the composite images), only the largest retinal vessels carrying the fastest flows are revealed in the retina. That would lead to a source of retro-illumination that would be very spatially localized, and the retinal structures would be much more visible in the pupil plane reconstruction. Using the 1-8 kHz (closer to the cyan part of the composite images) enables a much more uniform spatial distribution of energy in the fundus, and therefore to have less noticeable defocused retinal features in the pupil plane. Secondly, the effect of the cataract is simply more significant than the defocused vasculature. Also, we have here used mostly measurements where the optic nerve head (ONH) was imaged for the challenge of measuring the eye transparency at the same time as the ONH blood flow. However, imaging an area outside the ONH would be more favorable to having a homogeneous source of light from the retina. Finally, it should be mentioned that using a camera with smaller pixels is also expected to improve the defocusing of retinal features in the pupil plane, since smaller pixels enable a higher numerical aperture of the holographic detection, and therefore to reduce the depth of field of the reconstructed holograms~\cite{Kreis2006handbook}.



In Fig.~\ref{fig_7_surgery} are shown LDH measurements performed in an eye affected by cataract before (upper images) and after (lower images) the cataract surgery. The first column shows power Doppler images in the pupil plane, and the second and third columns show in the retinal plane the color composite images of slow/fast flows, and the blood flow movies obtained from the same data. Before the operation, the cataract is visible in the center of the pupil and on its edges (riders). However, images of acceptable quality are obtained in the retina.
After the operation, the retinal images become sharper in the area that corresponds to the projection of the cataract onto the retina. A part of the lower field of view has on the other hand become more blurred after the operation.
The arrow '1' indicates the limit of the capsulorhexis where the lens capsule was opened for the cataract extraction. The arrow '2' indicates the lower edge of the implant, the image is blurred below because of the absence of refractive implant.

\subsection{Corneal neovascularization}

\begin{figure}[t!]
\centering
\includegraphics[width = 1\linewidth]{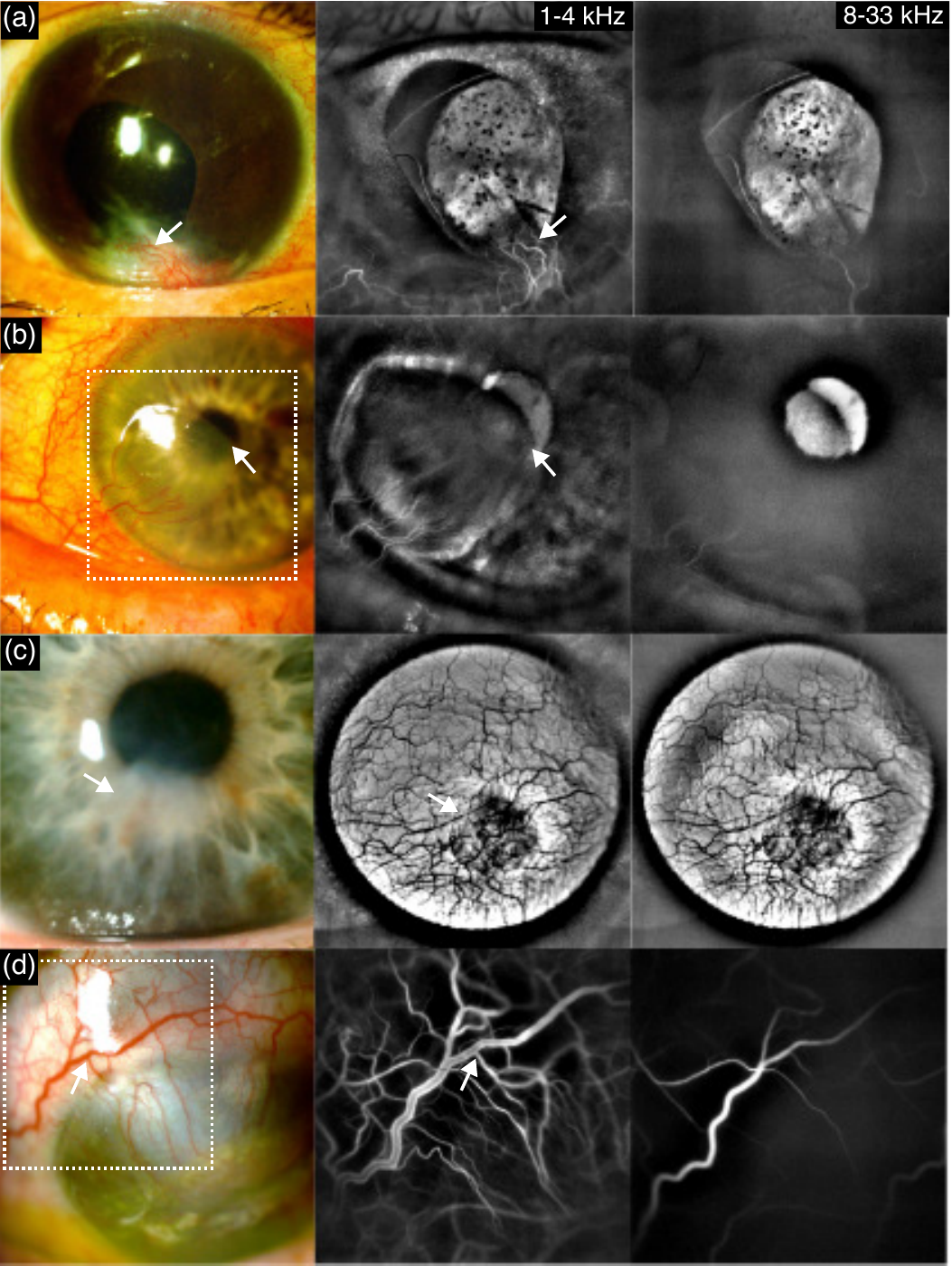}
\caption{LDH imaging of corneal neovascularization.
(a) Following a penetrating ocular trauma, \textcolor{blue}{\href{https://osapublishing.figshare.com/articles/media/Visualization_6_mp4/14224463}{Visualization 6}}.
(b) Corneal bacterial keratitis, \textcolor{blue}{\href{https://osapublishing.figshare.com/articles/media/Visualization_7_mp4/14224466}{Visualization 7}}.
(c) Case of stable corneal herpes.
(d) Corneal bacterial keratitis, \textcolor{blue}{\href{https://osapublishing.figshare.com/articles/media/Visualization_8_mp4/14224469}{Visualization 8}}.
}
\label{fig_6_Neovascular}
\end{figure}

Finally, we present in Fig.~\ref{fig_6_Neovascular} cases of corneal ulcers (keratitis) involving neovascularization imaged with LDH. For each example are shown the color photographs of the eye, and power Doppler images of the anterior segment for the low (1-4 kHz) and high frequency ranges (8-33 kHz). Because slit-lamp and LDH images are differently contrasted and because the pupil is more dilated in LDH images (LDH measurements are performed in the dark), the comparison is not easy for all cases. The arrows point to common features of the images to facilitate the comparison.
The first example features a case of corneal neovascularization following a penetrating ocular trauma. Neovascular corneal vessels have formed on the lower part of the eye, and some deposits on the posterior chamber implant can be observed.
The second example shows a case of corneal neovascularization following corneal bacterial keratitis. The corneal abscess induces a disorganization of the architecture of the corneal collagen fibers and therefore a loss of corneal transparency. The arrow points to the limit of the corneal scarring and it is interesting to note that the non-transparent area is seen with a darker contrast. This can be explained by the fact that in the areas where light is diffused in all directions, less light reaches the camera, and therefore the power Doppler in the area is lower and it appears darker.
In the power Doppler movies of these first two examples shown in \textcolor{blue}{\href{https://osapublishing.figshare.com/articles/media/Visualization_6_mp4/14224463}{Visualization 6}} and \textcolor{blue}{\href{https://osapublishing.figshare.com/articles/media/Visualization_7_mp4/14224466}{Visualization 7}}, white bursts due to bulk motions can be seen over time. For a reason unknown to us, these bursts seem brighter in the iris than in the surrounding conjunctiva.
The third example shows a case where the corneal neovascularization has completely covered the pupil, although it is not visible on the color photograph. The pupil is more dilated on the power Doppler images than on the color photograph, the arrow points to a non-transparent area visible with a bright contrast, that also appears on power Doppler images but with a dark contrast. Blood vessels are visible with a dark contrast in both frequency ranges, which is something we have never observed neither on fundus nor corneal images. This dark contrast of blood vessels means that blood flow is either circulating very slowly or not at all in these neovessels.
It could also be assumed that blood flow in these vessels is circulating very fast~\cite{Puyo2020b}, but that seems less plausible since they are of very small calibers.
Finally, the fourth example shows a case of corneal neovascularization following corneal bacterial keratitis. Vessels of large size can be seen on the color photograph, and these vessels can be seen in the high frequency power Doppler image, meaning that flow of relatively high speed circulates in these vessels.



\section{Discussion and conclusion}

In this manuscript, we have explored the potential clinical uses of LDH in the anterior segment. We first showed that LDH can be used to image blood flow with a high temporal resolution and over a field of view that can be easily adapted in the conjunctiva and episclera as well as in corneal neovascularization. The functional blood flow contrast brought by LDH such as the systole/diastole or slow/fast flows could be helpful to understand these vasculatures. In the case of neovascularization or inflammation, LDH could be helpful in monitoring and better understanding the developments of the pathology by providing at least semi-quantitative flow measurements and identify the vessels that carry arterial, venous, and potentially lymphatic flows. In the corneal neovascularizations we presented in this manuscript, it could be seen that neovessels were differently contrasted for different frequency ranges, which reflect the different flow velocities. In one case there was seemingly no flow circulating in the neovessels since the vessels were seen with a dark contrast at the lowest frequencies LDH can access. Therefore LDH presents a good potential to follow-up the evolution of neovacularization as it would bring a functional information complementary to the structural information that can be obtained with established angiographic instruments such as fluorescein and indocyanine green angiography~\cite{Nieuwenhuizen2003, Meyer1987low, Chan2001vascular, Marvasti2016, Aydin2000anterior, Alsagoff2001}, and optical coherence tomography angiography~\cite{Akagi2018, Schuerch2020, Hayek2019, Liu2019quantitative, Di2019optical}.
Despite the likely higher clinical interest of iris blood flow compared to conjunctival blood flow, we did not present images of iris blood flow in this article since images of iris vessels were at this point not obtained in all subjects. The iris vasculature appeared more challenging to reveal with LDH than the conjunctival one. Several parameters seemed to come into play. A smaller field of view seemed favorable, presumably because it allows increasing the illumination count per area. A smaller field of view is also expected to be beneficial to the SVD filtering. The strength of the SVD is to take advantage of the spatial coherence of bulk motion, but this spatial coherence is compromised when imaging a larger of view. Indeed, for a large of view, the direction of the motion of exposed tissues with respect to the illumination beam varies significantly throughout the field of view, which prevents the Doppler contribution of global movements to be isolated in the first few singular vectors. This is of high importance since the iris flow seemed to be revealed with frequencies of just a few kHz, which requires a SVD filtering since this frequency band completely overlaps with the Doppler response range of tissue motion. Another parameter that subjectively seemed of importance was the iris melanin pigmentation, as darker irises seemed to make the imaging of iris vessels more challenging. Finally, it seemed that constriction of the pupil induced by exposure to ambient light (miosis) was also favorable to image the iris vessels. Therefore, more work is needed to determine whether experimental conditions can be devised to enable a systematic imaging of the iris vasculature in all subjects.

We have also shown that thanks to the ability of digital holography to simultaneously image the pupil and retina, eye movements can be tracked with a good spatio-temporal resolution, roughly in the range of $\SI{1}{\milli\second}$ and about $\SI{5}{\micro\meter}$ in the retina, and $\SI{10}{\micro\meter}$ in the pupil. Retinal and pupil lateral displacements, the pupil diameter, and retinal blood flow could be jointly monitored from the same dataset. Information about the instantaneous axial velocity of bulk motion also seems accessible. There was a trade-off between spatial resolution and robustness: spatial resolution should be improved if imaging a smaller field of view, but that would make the registration less robust when the amplitude of eye movements becomes too large. The temporal resolution was here limited by the signal-to-noise ratio of power Doppler images, and therefore could probably be improved with a higher ocular irradiance and a higher camera frame rate. Although both the retinal and corneal irradiances were well under the maximum permissible exposure, we did not increase the ocular irradiance in order to stay within the limits established by our clinical trial. In contrast with off-axis holographic imaging, the sensitivity of on-axis holographic measurements is not a trivial point because of the parasitic contributions of the self-beating terms and of the twin-image~\cite{Hillmann2017}. However, it seems likely that the temporal resolution of the images would benefit from a higher irradiance. We also did not used the highest possible camera frame rate so as to record longer measurements since they are limited by the camera on-board memory. The ability to jointly measure the eye movements, pupil dilation/constriction, and blood flow variations is promising to study ocular biomechanics, investigate possible causal relationships between them, and potentially translate it into clinical applications~\cite{girard2015translating}. Finally, the ability to visualize the anterior part of the eye while imaging the fundus could also be of interest during the measurement or in post-processing to check if the subject pupil's position is satisfying.


Finally, we investigated the contrast that can be obtained with LDH in non-transparent corneas and lenses. We showed with cases of cataract that absorbing structures are revealed with a dark contrast, as would be dust on optical elements of the setup. We have observed with cases of corneal scarring that features causing diffusion also generates a clearly detectable contrast with LDH, as a loss of transparency leads to a signal loss. We found that we were able to perform LDH retinal blood flow examinations in the presence of mild cataract. We assume that this is due to the use of the full pupil aperture for both the illumination and collection of light from the eye fundus. Compared to slit lamp examinations, we found that LDH offers a better contrast of opacities, is not operator-dependent, highly repeatable, comfortable for the subject, and enables to document the anterior segment transparency over the whole pupil with a single image. A limitation of LDH is that the large depth of field of the reconstructed holograms does not allow to identify the axial position of the opaque structure as OCT is able to in the anterior segment~\cite{Grulkowski2018, AlbertDeCastro2018, Eugui2020, RodriguezAramendia2021}. LDH may also not be the most promising approach to image the cornea with cellular resolution~\cite{Cavanagh1993, Guthoff2009, Wartak2020, Auksorius2020cornea, Mazlin2020curved}. However, overall LDH shows great potential to perform an automated and objective assessment of the anterior segment transparency and could be useful to follow-up and document the evolution of cataracts. As a high temporal sampling frequency is not a requirement, transparency assessments of the anterior segment by LDH could be made relatively inexpensive. The eye tracking and transparency assessments presented in this article can be obtained from the usual LDH examinations that aim at imaging blood flow in the eye fundus. It simply requires positioning the eye backwards so that the pupil slightly limits the effective aperture, which we would otherwise try to avoid. However, as demonstrated here, retinal blood flow imaging can still be performed when trying to either track the eye movement or assess the anterior segment transparency. Another way of looking at the results in Fig.~\ref{fig_4_EyeMovements} and \ref{fig_5_Cataracte} is that LDH fundus blood flow examinations remain possible despite eye motion and mild cataract.

Our method of imaging the crystalline lens transparency bears resemblance to the technique presented by Weber and Mertz~\cite{Weber2020}, where the cornea and lens was imaged in transmission thanks to an asymmetric retro-illumination from the fundus. The authors also used a setup with a polarizing beamsplitter cube, but they used incoherent light and they restricted the imaged area to a 1 mm diagonal field of view. The authors demonstrated cellular resolution thanks to the high numerical aperture of their setup, and mention that the contrast brought by the forward-scattered light can help reveal some structures. The imaging of the transparency of the anterior segment we demonstrated with LDH relies on the numerical propagation of light backscattered by the fundus. We have here used the light decorrelated under the effect of blood flow, but it might also be feasible to use light reflected by tissue.


In conclusion, we have shown that the temporal fluctuations of monochromatic digital holograms of the eye recorded with a very basic optical setup can yield a wealth of information that enables to jointly perform ocular blood flow imaging, eye tracking, and the assessment of the lens and corneal transparency.

\section*{Funding Information}
European Research Council (Synergy HELMHOLTZ \#610110). Agence Nationale de la Recherche (FOReSIGHT ANR-18-IAHU-0001). Region Ile-de-France (EX047007 - SESAME 2019 - 4DEye).

\section*{Acknowledgments}
The authors would like to thank Dierck Hillmann for helpful discussion.

\section*{Disclosures}
The authors declare no conflicts of interest.

\section*{Data availability}
Data underlying the results presented in this paper are not publicly available at this time but may be obtained from the authors upon reasonable request.

\section*{Supplementary Material}
\noindent
\textcolor{blue}{\href{https://youtu.be/HIaXM8wVeiU}{Supplementary Visualization 1}}. \newline

\bibliography{./Bibliography}

\end{document}